\documentclass{article}

\usepackage{microtype}
\usepackage{graphicx}
\usepackage{subfigure}
\usepackage{booktabs} % for professional tables

\usepackage{hyperref}

\usepackage[accepted]{icml2018}

\icmltitlerunning{Machine learning-guided directed evolution for protein engineering}
\usepackage[superscript,biblabel]{cite}

\begin{document}

\onecolumn
%\twocolumn[
\title{Machine learning-guided directed evolution for protein engineering}
\maketitle



\icmlsetsymbol{equal}{*}

\begin{icmlauthorlist}
\icmlauthor{Kevin K. Yang}{cce}
\icmlauthor{Zachary Wu}{cce}
\icmlauthor{Frances H. Arnold}{cce}
\end{icmlauthorlist}

\icmlaffiliation{cce}{Division of Chemistry and Chemical Engineering, California Institute of Technology, Pasadena, CA, USA}

\icmlcorrespondingauthor{Frances H. Arnold}{frances@cheme.caltech.edu}

\icmlkeywords{Machine Learning, Protein Engineering, Directed Evolution}

\vskip 0.3in


\printAffiliationsAndNotice{}  % leave blank if no need to mention equal contribution

% ---------- abstract ------------
\begin{abstract}%   <- trailing '%' for backward compatibility of .sty file
Machine learning (ML)-guided directed evolution is a new paradigm for biological design that enables optimization of complex functions. ML methods use data to predict how sequence maps to function without requiring a detailed model of the underlying physics or biological pathways. To demonstrate ML-guided directed evolution, we introduce the steps required to build ML sequence-function models and use them to guide engineering, making recommendations at each stage. This review covers basic concepts relevant to using ML for protein engineering as well as the current literature and applications of this new engineering paradigm. ML methods accelerate directed evolution by learning from information contained in all measured variants and using that information to select sequences that are likely to be improved. We then provide two case studies that demonstrate the ML-guided directed evolution process. We also look to future opportunities where ML will enable discovery of new protein functions  and uncover the relationship between protein sequence and function.

\end{abstract}

%%% Local Variables:
%%% mode: latex
%%% TeX-master: "main"
%%% End:

% ---------- intro ------------
\section{Introduction}
Protein engineering seeks to design or discover proteins whose properties, useful for technological, scientific, or medical applications, have not been needed or optimized in nature. A protein's function, such as its expression level, catalytic activity, or other properties of interest to the protein engineer, is specified by its amino-acid sequence. Protein engineering inverts this relationship in order to find a sequence that performs a specified function. Unfortunately, current biophysical prediction methods cannot distinguish between the functional level of closely-related proteins, and we cannot reliably map sequence to function\cite{Dou2017designchallenges, garcia2017computational}. Furthermore, the space of possible protein sequences is too large to be searched exhaustively naturally, in the laboratory, or computationally\cite{mandecki1998game}. The problem of finding optimal sequences is NP-hard, meaning there is no known polynomial-time method for searching this space\cite{pierce2002protein}. Functional proteins are also extremely scarce in this vast space of sequences. Moreover, as the threshold level of function increases, the number of sequences having that function decreases exponentially\cite{smith1970natural,orr2006distribution}. As a result, highly functional sequences are vanishingly rare and overwhelmed by nonfunctional and mediocre sequences. 

% Until recently, the two main approaches for finding high-fitness protein sequences have been directed evolution and rational design. Rational design uses physics-based models to guide the search for improved sequences. These models typically contain an atomic structural representation of a protein and energy-based scoring functions to quantify the target function\cite{dahiyat1997novo,Siegel2010design}. Rational design has been successful in identifying sequences that fold into desired static structures\cite{dou2018novo}. This is an important advance, and useful when a single stable structure dictates function\cite{dou2018novo}. However, because proteins are marginally stable, even small inaccuracies in energy-based scoring functions can lead to very poor performance\cite{Dou2017designchallenges}. 

% Many protein functions such as binding, catalysis, allostery, and signalling, are mediated through recessed cavities, mobility, or multiple low-energy states\cite{huang2016coming}. For example, computational enzyme design currently proceeds by designing an idealized active site for the desired reaction, matching the active site residues to stable backbones, and then using molecular dynamics (MD) simulations to winnow designs with flaws not apparent from static evaluations. MD simulations require enormous computational resources (100s of CPU hours for each variant) and are not appropriate for testing many variants. This process generally yields sequences with modest activity that are finally improved with directed evolution\cite{garcia2017computational}. 

\begin{figure}
	\centering
	\includegraphics[width=\textwidth]{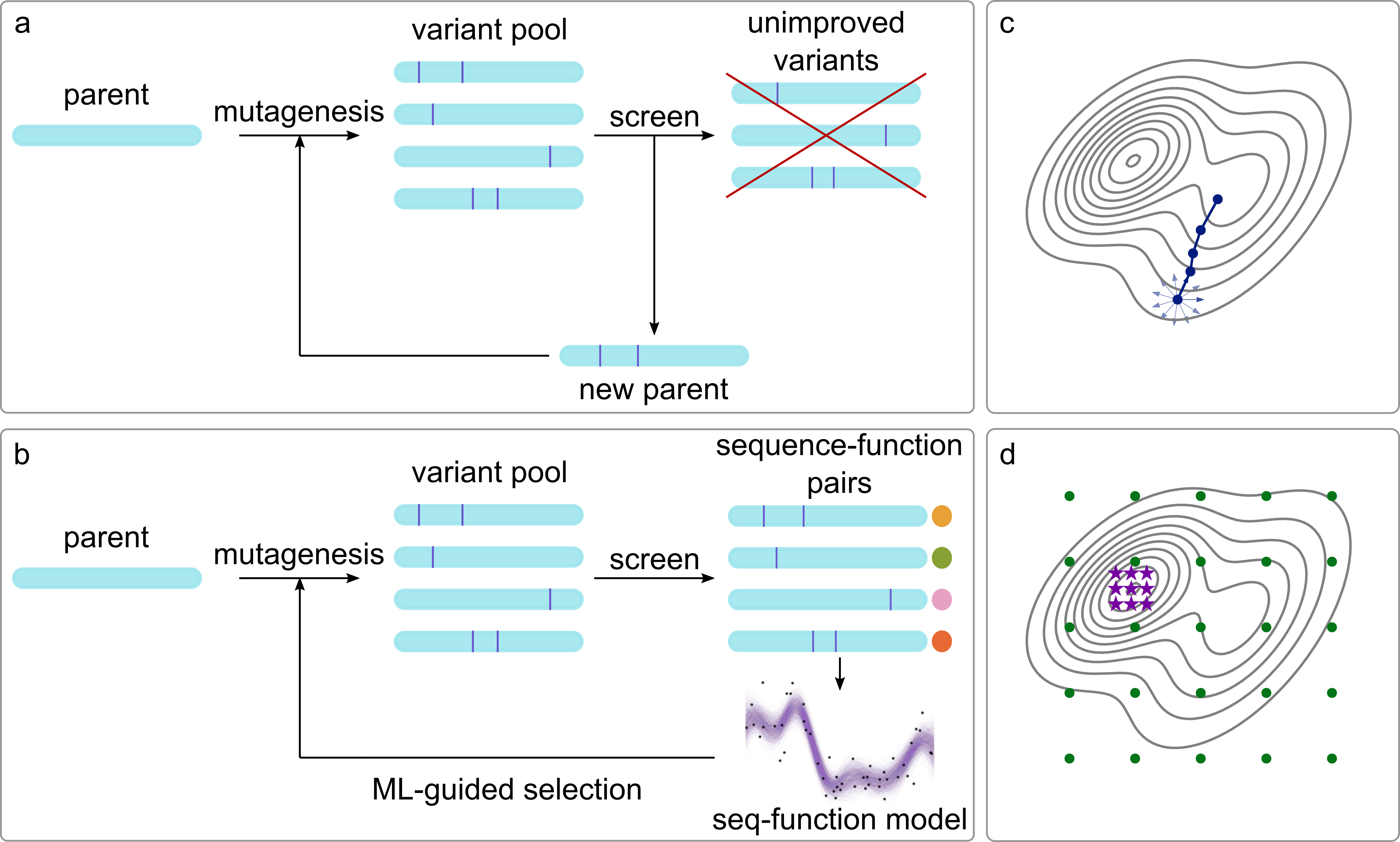}
	\caption{Directed evolution with and without machine learning. (a) Directed evolution uses iterative cycles of diversity generation and screening to find improved variants. Information from unimproved variants is discarded. (b) Machine-learning methods use the data collected in each round of directed evolution to choose the mutations to test in the next round. Careful choice of mutations to test decreases the screening burden and improves outcomes. (c) Directed evolution is a series of local searches on the function landscape. (d) Machine learning-guided directed evolution often rationally chooses the initial points (green circles) to maximize the information learned from the function landscape, allowing future iterations to quickly converge to improved sequences (violet stars).}
	\label{fig:mlde}
\end{figure}

Directed evolution has been successful because it sidesteps our inability to map protein sequence to function. Inspired by natural evolution, directed evolution climbs a fitness landscape by accumulating beneficial mutations in an iterative protocol of mutation and selection. As shown in Figure~\ref{fig:mlde}, the first step is sequence diversification using techniques such as random mutagenesis, site-saturation mutagenesis, or recombination to generate a library of modified sequences starting from the parent sequence(s). In the second step, screening or selection identifies variants with improved properties for the next round of diversification. These steps are repeated until fitness goals are achieved. Illustrated in Figure~\ref{fig:mlde}c, directed evolution finds local optima through repeated local searches, taking advantage of functional promiscuity and the fact that functional sequences are clustered in sequence space\cite{khersonsky2010enzyme, smith1970natural}. 

Directed evolution is limited by the fact that even the most high-throughput screening or selection methods only sample a fraction of the sequences that can be made using most diversification methods, and developing efficient screens is nontrivial. There are an enormous number of ways to mutate any given protein: for a 300-amino acid protein there are 5,700 possible single amino acid substitutions and 32,381,700 ways to make just two substitutions with the 20 canonical amino acids. Screening exhaustively to find rare beneficial mutations is expensive and time-consuming and sometimes simply impossible. Moreover, directed evolution requires at least one minimally-functional parent and a locally-smooth sequence-function landscape for stepwise optimization\cite{romero2009exploring}. Recombination methods may allow for bigger jumps in sequence space while retaining function\cite{drummond2005conservative}, but sequences designed using recombination are by definition restricted to exploring combinations of previously-explored mutations. No matter the diversification technique, directed evolution is energy-, time-, and material-intensive, and multiple generations may be required to achieve meaningful performance improvements.

While directed evolution discards information from unimproved sequences, machine-learning methods can use this information to expedite evolution and expand the properties that can be optimized by intelligently selecting new variants to screen, reaching higher fitnesses than through directed evolution alone\cite{wu2019combination}. Figure~\ref{fig:mlde}b illustrates this data-augmented cycle. Machine-learning methods learn functional relationships from data\cite{hastiefriedman,murphy2012machine_note} -- the only added costs are in computation and DNA sequencing, the costs of which are decreasing rapidly. Machine-learning models of protein function can be predictive even when the underlying mechanisms are not well-understood. Furthermore, and perhaps most importantly, machine-learning guided directed evolution is able to escape local optima by learning efficiently about the entire function landscape, as illustrated in Figure~\ref{fig:mlde}d.

Machine learning is not necessarily useful in all applications. Because one major benefit of machine learning is in reducing the quantity of sequences to test, machine learning will be particularly useful in cases where lack of a high-throughput screen limits or precludes directed evolution. However, when a sufficient number of sequences can be screened or if the fitness landscape is smooth and additive, machine learning may not significantly decrease screening burden or find better variants. In these cases, the added cost of sequencing DNA to form sequence-function relationships is unnecessary.  

Once the decision has been made to use machine learning, there are two key steps: i) building a sequence-function model and ii) using that model to choose sequences to screen. We provide practical guides for these steps as well as two case studies that illustrate the machine learning-guided directed evolution process. Finally, we consider developments that will enable wider applications of machine learning for protein engineering.
% ---------- machine learning and directed evolution ------------
% \section{Machine learning-guided directed evolution}

% In most computational methods, the user provides a hard-coded algorithm and inputs, and the computer executes the steps provided by the human expert. In contrast, machine-learning models infer patterns from data, which can then be used to make predictions on unobserved data.  The trained model can then be used to guide evolution.

\section{Building a machine-learning sequence-function model}

Machine-learning models require examples of protein sequences and their resulting functional measurement in order to learn. The initial sequences chosen determine what the model can learn. The initial set of variants to screen can be selected in different ways: i) at random from the library\cite{fox2007improving}, ii) to maximize information about the mutations considered\cite{liao2007engineering,govindarajan2014mapping,musdal2017exploring}, or iii) to maximize information about the remainder of the library\cite{romero2013navigating_note,bedbrook2017machine,bedbrook2017minimally}. Selecting variants at random is usually the simplest method; however, for low-throughput screens, it can be very important to maximize information obtained from high-cost experiments. Maximizing information about mutations is roughly equal to seeing each mutation considered in as many contexts as possible. Maximizing information about the remainder of the library is roughly equal to maximizing diversity in the training sequences. After collecting the initial training data, the user must decide what type of machine-learning model to use, represent the data in a form amenable to the model, and train the model.

\subsection{Choosing a model}

A wide range of machine-learning algorithms exist, and no single algorithm is optimal across all learning tasks\cite{wolpert1996lack}. For machine learning-guided directed evolution, we are most interested in methods that take input sequences and their associated output values and learn to predict the outputs of unseen sequences. 

The simplest of these machine-learning models apply a linear transformation of the input features, such as the amino acid at each position, the presence or absence of a mutation\cite{fox2007improving}, or blocks of sequence in a library of chimeric proteins made by recombination\cite{li2007diverse}. Linear models are commonly used as baseline predictors before more powerful models are tried. 

Classification and regression trees\cite{breiman2017classification} use a decision tree to go from input features (represented as branches) to labels (represented as leaves). Decision tree models are often used in ensemble methods, such as random forests\cite{breiman2001random} or boosted trees\cite{friedman2002stochastic}, which combine multiple models into a more accurate meta-predictor. For small biological datasets ($<10^4$ training examples), including those often encountered in protein engineering experiments, random forests are a strong and computationally efficient baseline; these have been used to predict enzyme thermostability\cite{tian2010predicting,li2012prots,jia2015structure}.

Kernel methods, such as support vector machines\cite{cortes1995support} and kernel ridge regression\cite{nadaraya1964krr}, employ a kernel function, which calculates similarities between pairs of inputs, to implicitly project the input features into a high-dimensional feature space without explicitly calculating the coordinates in this new space. While general-purpose kernels can be applied to protein inputs, there are also kernels designed for use on proteins, including spectrum and mismatch string kernels\cite{leslie2001spectrum,leslie2004mismatch}, which count the number of shared subsequences between two proteins, and weighted decomposition kernels\cite{jokinen2018mgpfusion}, which account for three-dimensional protein structure. Support vector machines have been used to predict protein thermostability\cite{capriotti2005mutant2,capriotti2005predicting,cheng2006prediction,buske2009silico, tian2010predicting,jia2015structure,li2012prots,liu2012grading}, enzyme enantioselectivity\cite{zaugg2017learning}, and membrane protein expression and localization\cite{saladi2018statistical}.

Gaussian process models combine kernel methods with Bayesian learning to produce probabilistic predictions\cite{rasmussen2006gaussian}. These models rigorously capture uncertainty, which can provide principled ways to guide experimental design in optimizing protein properties. The run-time for exact GP regression scales as the number of training examples cubed, making it unsuitable for large ($>10^3$) datasets, but there are now fast and accurate approximations\cite{wilson2015kernel,wang2019exact}. Gaussian processes have been used to predict thermostability\cite{pires2013mcsm,romero2013navigating_note,jokinen2018mgpfusion}, substrates for enzymatic reactions\cite{mellor2016semisupervised}, fluorescence\cite{saito2018machine}, membrane localization\cite{bedbrook2017machine},  and channelrhodopsin photo-properties\cite{bedbrook2017minimally}.

Deep learning models, also known as neural networks, stack multiple linear layers connected by non-linear activation functions, allowing them to extract high-level features from structured inputs. Neural networks are well-suited for tasks with large labeled datasets with examples from many protein families, such as protein-nucleic acid binding\cite{zhang2015deep,alipanahi2015predicting,zeng2016convolutional}, protein-MHC binding\cite{hu2017deepmhc}, binding site prediction\cite{jimenez2017deepsite}, protein-ligand binding\cite{gomes2017atomic,mazzaferro2017predicting}, solubility\cite{khurana2018deepsol}, thermostability\cite{dehouck2009fast, giollo2014neemo}, subcellular localization\cite{almagro2017deeploc}, secondary structure\cite{sonderby2014protein}, functional class\cite{szalkai2018near,cao2017prolango}, and even 3D structure\cite{Hopf2012membrane}. 

\begin{figure}
	\centering
	\includegraphics[width=0.8\textwidth]{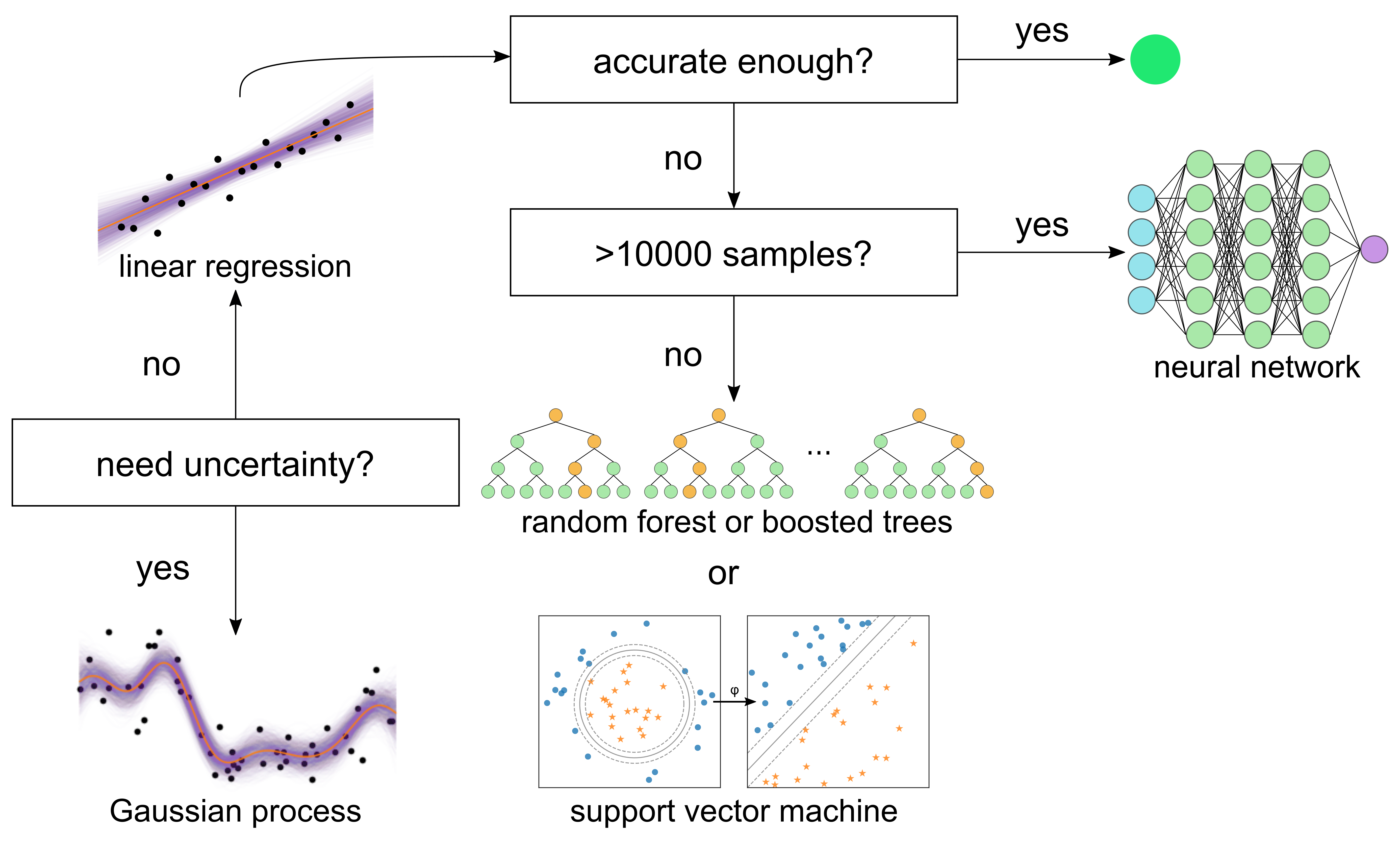}
	\caption{A general heuristic for choosing a machine-learning sequence-function model for proteins.}
	\label{fig:ml}
\end{figure}

Figure~\ref{fig:ml} shows a general heuristic for choosing an algorithm for modeling protein sequence-function relationships. If estimates of model uncertainty are required, Gaussian processes are the simplest off-the-shelf model family. Otherwise, linear models provide a simple baseline to more complex models. If a linear model is insufficiently accurate, random forests, boosted trees, or support vector machines are fast and efficient for datasets with fewer than 10,000 examples, while neural networks generally provide the best performance on larger datasets.

\subsection{Model training and evaluation}

Training a machine learning model refers to tuning its parameters in order to maximize predictive accuracy. The key test for a machine-learning model is the ability to accurately predict labels for inputs it has not seen during training. Therefore, when training models, it is necessary to estimate the model's performance on data ~\textit{not} in the training set. Thus it is essential to remove a portion of the data, called the test set, until the absolute end for model evaluation. Typically, the test set comprises approximately 20\% of the data. 

In addition to parameters, all model families have hyperparameters that determine the form of the model. In fact, the family of model chosen is itself a hyperparameter. Unlike model parameters, hyperparameters cannot be learned directly from the data. These may be set manually by the practitioner or determined using a procedure such as grid search, simulated annealing, random search, or Bayesian optimization\cite{snoek2012practical}. For example, in support vector machines, the type of kernel is a hyperparameter, as are the number of layers and learning rate in a deep neural network. The vectorization method is also a hyperparameter. Even modest changes in the values of hyperparameters can improve or diminish accuracy considerably, and the selection of optimal values is often challenging, as each set of hyperparameters considered may require training a new version of the model. 

In order to compare models and select hyperparameters during a study, the data that remain after the test set has been removed should be split into a training set and a validation set. The training set is used to learn model parameters, while the validation set is used to choose between models with different hyperparameters by providing an estimate of the test error. If the training set is small, cross-validation may be used instead of a constant validation set. In $n$-fold cross-validation, the training set is partitioned into $n$ complementary subsets. Each subset is then predicted using a model trained on the remaining subsets. Averaging accuracy across the withheld subsets provides an estimate of predictive accuracy over the entire training set. Cross-validation provides a better estimate of the test error than using a constant validation set but requires more training time. 

Care must be taken when selecting the training/validation/test sets that the splits allow an accurate estimate of model performance under the conditions where it will be used. Datasets from mutagenesis studies tend to be small and focused. In this case, the best practice is to train on variants characterized in earlier rounds of mutagenesis and to evaluate model performance on later rounds in order to recapitulate the iterative engineering process. When dealing with large, diverse datasets containing examples from different protein families, the best practice is to ensure that all examples in the validation and test sets are some minimum distance away from all the training examples in order to test the model's ability to generalize to unrelated sequences.

\subsection{Vector representations of proteins}
Machine-learning models act on vectors or matrices of numbers, so protein sequences must be vectorized before model training. How each protein sequence is represented determines what can be learned\cite{domingos2012few,bengio2013representation}. In general, a protein sequence is a string of length $L$ where each residue is chosen from an alphabet of size $A$. The simplest way to encode such a string is to represent each of the $A$ amino acids as a number. However, the assignment of each residue to a number enforces an ordering on the amino acids that has no physical or biological basis. Instead of representing each position as a single number, a one-hot encoding represents each of the $L$ positions as $A - 1$ zeros and one 1, with the position of the 1 within the series denoting the identity of the amino acid at that position. Given structural information, the identity of pairs of amino acids within a certain distance in the structure can also be one-hot encoded\cite{romero2013navigating_note,bedbrook2017machine}. Single mutations can also be encoded as a 20-dimensional vector where the original amino acid is denoted by -1, the new amino acid by 1, and all others by zero\cite{capriotti2004neural}. One-hot encodings are inherently sparse, memory-inefficient, and high-dimensional. In a one-hot encoding, there is no notion of similarity between sequence or structural elements: they are either identical or not. Nevertheless, one-hot encodings offer good performance for little complexity and can be considered a good baseline encoding. 

A protein can also be encoded by its physical properties by representing each individual amino acid with a collection of physical properties, such as its charge or hydrophobicity, and each protein with a combination of those properties. Properties such as predicted secondary structures can also be used to represent proteins. However, there are a large number of physical properties that could be used to describe each amino acid or protein, and the molecular properties that dictate each functional property are unknown. AAIndex\cite{kawashima2007aaindex} and ProFET\cite{ofer2015profet} attempt to systematically collect descriptors of protein sequences. There have also been attempts to describe each amino acid using two reduced dimensions based on volume and hydrophobicity\cite{barley2018improved} and to combine physical properties with structural information\cite{qiu2007structural,buske2009silico} by encoding each position in the sequence as a combination of the properties of amino acids in its geometric neighborhood. 

While a large number of protein sequences have been deposited in databases, only a tiny fraction are labeled with measured properties relevant to any specific prediction task. The unlabeled sequences contain information about the frequency and patterns of amino acids selected by evolution to compose proteins, information that may be helpful across prediction tasks. The simplest examples incorporating evolutionary information are BLOSUM\cite{henikoff1992amino} or AAIndex2-style substitution matrices based on relative amino-acid frequencies. However, more sophisticated continuous vector encodings of sequences can be learned from patterns in unlabeled sequences\cite{asgari2015continuous, mazzaferro2017predicting, ng2017dna2vec, kimothi2016distributed, yang2018learned,schwartz2018deep,alley2019unified} or from structural information\cite{bepler2019learning}. These representations learn to place similar sequences close together in the continuous space of proteins. Learned encodings are low-dimensional and may improve performance by transferring information in unlabeled sequences to specific prediction tasks. However, it is difficult to predict which learned encoding will perform well for any given task. 

Just as no model will be optimal across all machine learning tasks, there is no universally optimal vectorization method\cite{wolpert1996lack}. Researchers must use a combination of domain expertise and heuristics to select a set of encodings to compare. For small datasets, one-hot encodings offer superior performance to general sets of protein properties\cite{yang2018learned}, although careful feature selection informed by domain knowledge may yield more accurate predictions. If accuracy is insufficient, learned encodings may be able to improve performance. The encoding should ultimately be chosen empirically to maximize predictive performance.

\section{Using sequence-function predictions to guide exploration}

Once a sequence-function model has been trained, the next set of sequences to screen can be chosen either by collecting mutations believed to be beneficial or by directly optimizing over sequences. To label mutations as beneficial, linear models of the mutational effects can be learned and the parameters can be directly interpreted to classify mutations as beneficial, neutral, or deleterious. The most beneficial mutations can then be fixed, deleterious mutations can be eliminated from the pool of considered mutations, and new mutations can be added to the pool\cite{fox2007improving}. Alternately, the model can be used to select combinations of mutations that have a high probability of improving function\cite{saito2018machine} or to directly predict highly improved variants \cite{wu2019combination}.

\begin{figure}
	\centering
	\includegraphics[width=1.0\textwidth]{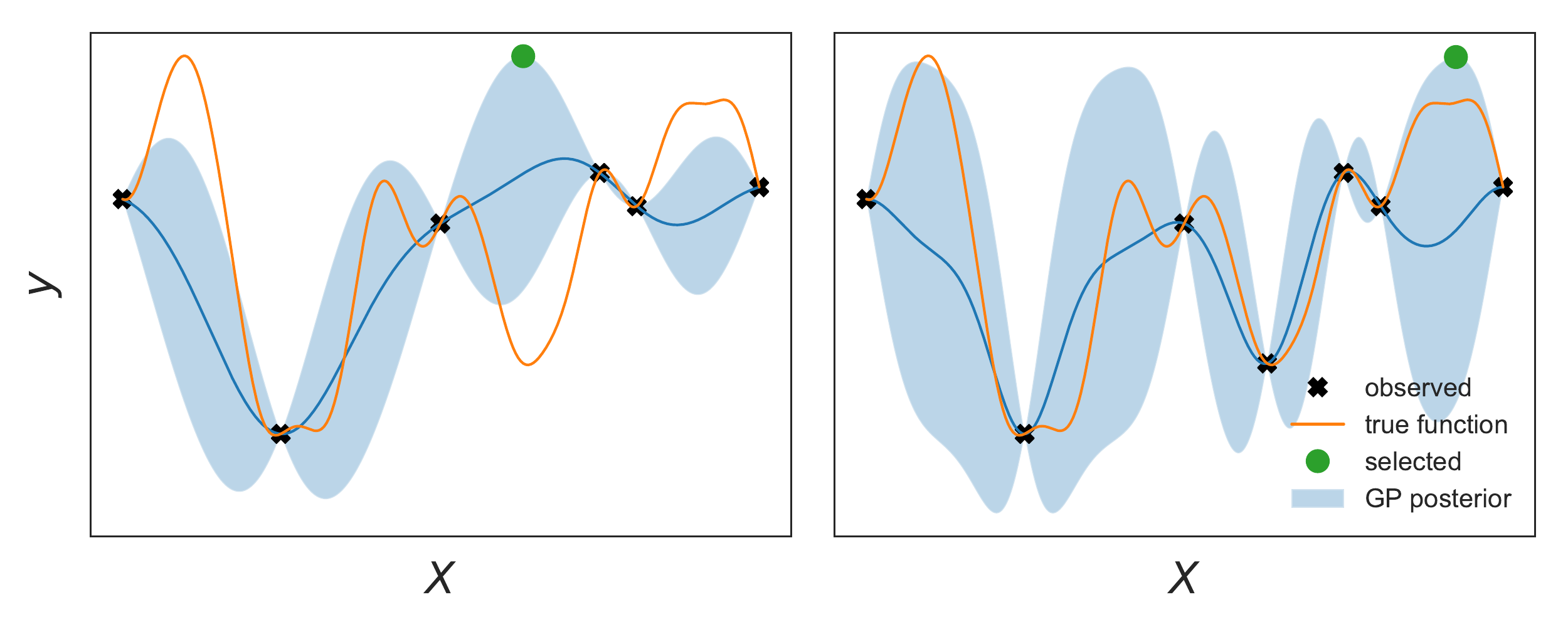}
	\caption{Gaussian Process Upper Confidence Bound algorithm. At each iteration, the next point to be sampled is chosen by maximizing the weighted sum of the posterior mean and standard deviation. This balances exploration and exploitation by exploring points that are both uncertain and have a high posterior mean. The right panel shows the posterior mean and standard deviation after observing the selected point (orange) in the left panel.}
	\label{fig:ucb}
\end{figure}

Learning and selection can also be performed directly over sequences. This can be as simple as enumerating all the sequences considered, using the trained model to predict their function, and then synthesizing the best predicted variants. However, if multiple rounds of engineering are to be performed and the sequence-function model provides probabilistic predictions, Bayesian optimization provides a principled way to trade off between exploiting the information learned from previous iterations and exploring unseen regions of sequence space at each iteration~\cite{snoek2012practical}. Probabilistic predictions provide a well-calibrated measure of uncertainty in addition to predicting an expected value: the model knows what it does not know. For example, the Gaussian Process Upper Confidence Bound (GP-UCB) algorithm balances exploration and exploitation by selecting variants that maximize a weighted sum of the predictive mean and standard deviation\cite{srinivas2009gaussian}, and is guaranteed to asymptotically minimize the cumulative regret (difference between sampled variants and best variant) over infinite iterations. Figure~\ref{fig:ucb} demonstrates two iterations of the GP-UCB algorithm. Alternatively, the model and data can be fully exploited using the Gaussian Process Lower Confidence Bound algorithm, which selects variants that maximize the weighted difference between the predictive mean and standard deviation. These approaches have been combined with structure-guided recombination to optimize cytochrome P450 thermostability\cite{romero2013navigating_note}, channelrhodopsin localization to mammalian cell membranes\cite{bedbrook2017machine}, and channelrhodopsin light-activated conductance\cite{bedbrook2017minimally}. Because there is no high-throughput screen for the channelrhodopsin properties, it would not have been possible to optimize conductance by traditional directed evolution.

% ---------- case studies ------------------
\section{Case studies}

\subsection{Using partial least squares regression to maximize enzyme productivity}
\begin{figure}
	\centering
	\includegraphics[width=\textwidth]{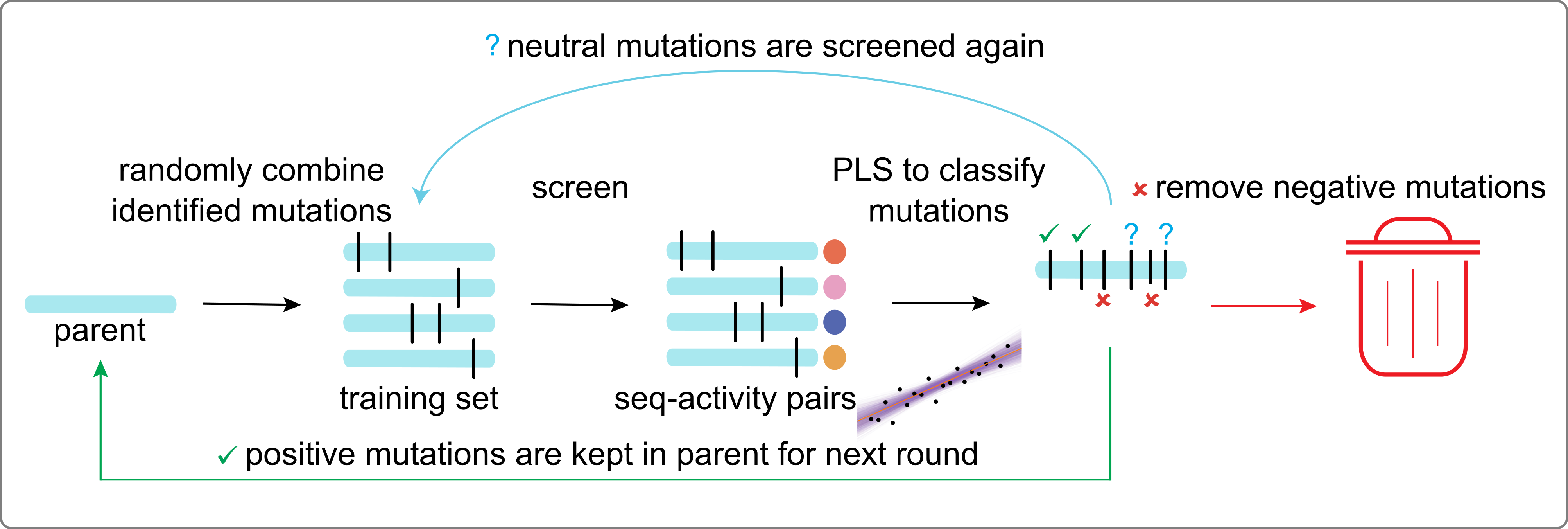}
	\caption{Directed evolution using partial least squares (PLS) regression. In this approach, Fox \textit{et al.} randomly recombine mutations previously identified through classical techniques such as random or site-directed mutagenesis\cite{fox2007improving}. Variants with these mutations are screened and sequenced, and the data are used to fit a linear model with the PLS algorithm. Based on the magnitude and sign of the contributions of the linear model, mutations are classified as beneficial, neutral, or deleterious, after which mutations are fixed, retested, or removed, respectively. With this approach, Fox \textit{et al.} were able to improve the volumetric productivity of a protein-catalyzed cyanation reaction roughly 4000-fold in 18 rounds of evolution.}
	\label{fig:prosar}
\end{figure}

An early large-scale evolution campaign guided by machine learning was performed by Fox \textit{et al.} in 2007 to improve the volumetric productivity of a halohydrin dehalogenase in a cyanation reaction by roughly 4000-fold\cite{fox2007improving}. In each round of evolution, summarized in Figure~\ref{fig:prosar}, 10-30 mutations of interest were first identified through random mutagenesis and site-saturation mutagenesis (including sites identified by tertiary structure analysis or sequence homology) and screening. These mutations were then randomly recombined, from which a number of variants equal to three times the number of positions mutated were sequenced and represented as one-hot vectors to form the input sequences used to train their model of choice, partial least squares (PLS; also known as projection to latent structures) regression\cite{fox2003optimizing_note}.  The PLS algorithm projects both sequences and fitnesses to a space with reduced dimension to fit the linear model. Thus it is able to fit data where the number of variables exceeds the number of observations and potentially avoids indirect correlations in the model\cite{dejong1993PLS}. The resulting linear model can be expressed  as
% What is $N$? -- removed
% Explanation of PLS seems a little too sparse. How is it different from regular linear regression? 

\begin{equation}
y = \sum_{m = 1}^{q} c_{m}x_{m}
\end{equation}

\noindent where $c_m$ is the additive contribution of each mutation to fitness, and $x_m$ indicates the presence ($x_m = 1$) or absence ($x_m = 0$) of the mutation. In this form, the PLS model has the distinct advantage of being readily interpreted. The authors classified each mutation as either beneficial, deleterious, or neutral based on the magnitude and sign of each coefficient. The classification of the mutation determines whether the mutation is retained, discarded, or tested again in the next round of evolution. However, when the Pearson's $r$ from leave-one-out cross-validation was low, the authors did not have much confidence in the model and would be more biased toward keeping mutations to test in future rounds. Finally, the best variant identified in the library with randomly recombined mutations was fixed as the starting sequence for the next round.

This case study was one of the first large protein engineering campaigns guided by statistical analysis. A total of 519,045 variants were tested, 268,624 from the initial diversity generation used to identify mutations to model with PLS and 250,421 from libraries designed by their approach. In 18 rounds of evolution, no variant showed more than a threefold improvement in any individual round. The evolution likely could have been accelerated by testing different vectorization methods and machine-learning algorithms to improve the accuracy at each round. The linear form of the model was used based on the assumption that local regions of the sequence-function landscape display predominantly additive effects. For fitness landscapes where this is not the case, an alternative model must be used. This work, which followed validation of the approach on a theoretical fitness landscape\cite{fox2003optimizing_note}, remains a landmark effort in the application of statistical modeling to protein engineering.

\subsection{Using Bayesian optimization to maximize thermostability of a cytochrome P450}

Romero \textit{et al.}\cite{romero2013navigating_note} demonstrate the utility of a machine-learning method that is particularly suitable in cases where it is expensive or difficult to screen in a directed evolution experiment. These authors sought to increase the thermostability, as measured by the $T_{50}$, of cytochrome P450s generated by recombining sequence fragments from the heme domains of the bacterial cytochrome P450 enzymes CYP102A1, CYP102A2, and CYP102A3. $T_{50}$ is defined as the temperature at which an enzyme loses half its activity after a 10-minute incubation. The sequence fragments were chosen to minimize the number of contacts broken, where contacts are amino acids within a certain distance (4.5 \AA) of each other. Chimeric genes can be made by directly synthesizing the DNA sequence for each construct, but this is expensive for large numbers of sequences. Furthermore, the $T_{50}$ measurement requires multiple incubations and measurements for each variant and is not particularly high-throughput. Therefore, this was a case where engineering without machine learning would have been difficult. 

The authors trained initial Gaussian process models for $T_{50}$ and the presence or absence of function on 242 chimeric P450s, and then validated model performance on a test set of chimeric P450s generated with different boundaries between sequence fragments. The Gaussian process model used a one-hot representation of the protein's 3-dimensional structure, which was demonstrated to be more predictive than a one-hot representation of the primary sequence. Gaussian process models are a good fit for this problem setting where only a small amount of data were available. These models also provide probabilistic predictions, which can be used to guide data-efficient exploration and optimization, but the computational requirements scale poorly with the number of training examples. 

\begin{figure}
	\centering
	\includegraphics[width=\textwidth]{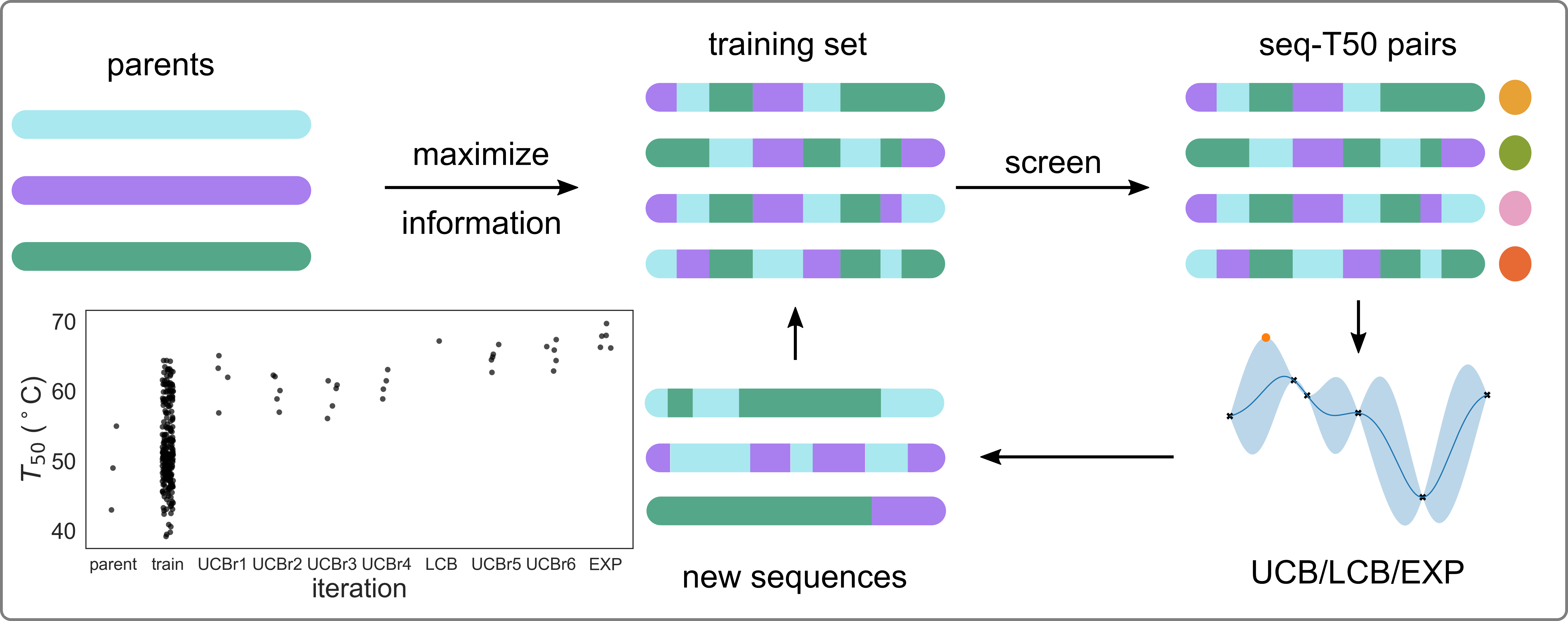}
	\caption{Directed evolution using Gaussian processes and Bayesian optimization. After an initial training set chosen to be maximally informative, subsequent batches of sequences are chosen using the Gaussian process upper confidence bound (UCB) or lower confidence bound (LCB) algorithms, or to fully exploit the model (EXP). The inset shows the $T_{50}$ for variants found in each round.}
	\label{fig:romero}
\end{figure}

After validating the accuracy of their models, the authors wanted to select a small set of additional sequences to bolster their models' knowledge of the recombination landscape. In general, this can be done by selecting those sequences that most reduce uncertainty in the predictions. This is measured by the mutual information between the measured sequences and the remaining sequences. Typically, these sequences will be very diverse. However, because many variants are non-functional and therefore provide no information about $T_{50}$, Romero \textit{et al.} used their classification model to select 26 additional sequences by maximizing the expected mutual information\cite{romero2013navigating_note}. Informally, these 30 sequences combined a high probability of being functional with high sequence diversity, and in fact, 26 of these sequences were functional despite being on average 106 mutations from the closest parent. Thus, this demonstrates the ability of a machine learning model to efficiently explore very diverse chimeric sequences while minimizing the resources wasted on screening non-functional proteins. 

With sufficient training data collected, Bayesian optimization was then used to search for more thermostable variants. First, four rounds of the batch Gaussian process upper-bound algorithm yielded a diverse sampling of thermostable P450s. However, because none of these variants increased the maximum observed $T_{50}$, the authors checked their sequence-function model by screening a sequence predicted to be stabilized with high certainty. Two additional iterations of GP-UCB were then followed by a pure exploitation round of five sequences, two of which were more thermostable than any previously observed P450s. 

By using previously collected data, an accurate sequence-function model, and Bayesian optimization, the authors thus demonstrated a framework for data-efficient protein engineering, which has since been transferred to other protein systems and properties\cite{bedbrook2018learning, bedbrook2017machine}. The framework and results are summarized in Figure~\ref{fig:romero}. 
% ---------- conclusions ------------
\section{Conclusions and future directions}

Supervised machine-learning methods have demonstrated their utility in directed protein evolution. However, broader applications of machine learning will require taking advantage of unlabeled protein sequences or sequences labeled for properties other than those of specific interest to the protein engineer. Databases such as UniProt\cite{uniprot2016uniprot} contain hundreds of millions of protein sequences, some of which are annotated with structural or functional information. These sequences contain information about the sequence motifs and patterns that result in a functional protein, and the structural and functional annotations provide clues as to how structure and function arise from sequence. These annotations can be learned from sequences\cite{schwartz2018deep}, and embeddings trained on these annotations may be able to transfer knowledge from UniProt to specific problems of interest\cite{pan2010survey}. 

These large quantities of unlabeled or partially-labeled sequence data may also enable machine-learning models to generate artificial protein diversity leading to novel protein functions. Only a tiny fraction of the amino acid landscape encodes functional proteins, and the complete landscape is plastered with cliffs and holes, where small changes in sequence result in complete loss of function. Natural and designed proteins are samples from the distribution of functional proteins, although these samples are biased by nature's evolutionary methods for generating proteins. A method for selectively sampling from the distribution of functional proteins would enable large jumps to previously unexplored sections of sequence space that may contain novel functions. Generative models of the distribution of functional proteins provide such a tool, and are an attractive alternative to~\textit{de novo} design methods\cite{baker2010exciting}. 
% Worth talking about nature's bias? "Natural and designed proteins are samples from the distribution of functional proteins, although these samples are biased by nature's evolutionary methods for generating proteins. 

\begin{figure}
	\centering
	\includegraphics[width=0.7\textwidth]{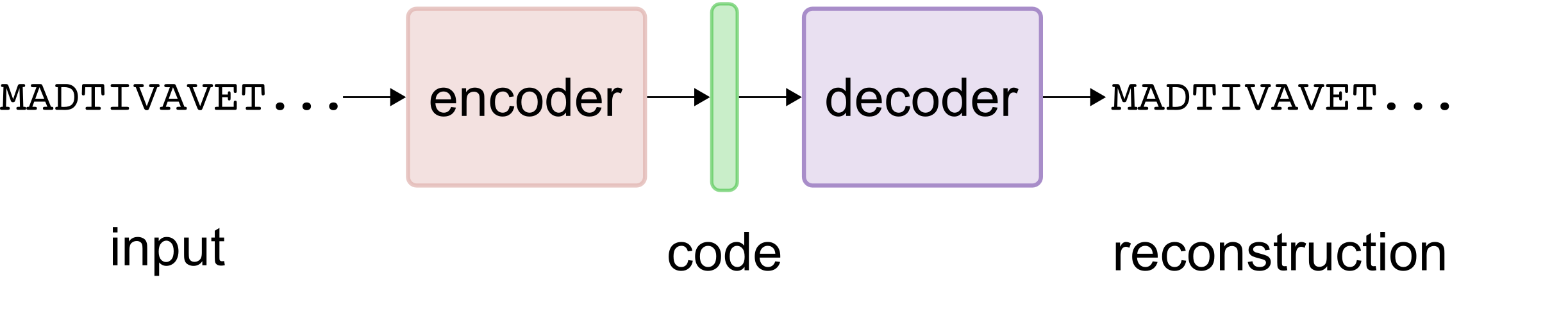}
	\caption{Autoencoder. An autoencoder consists of an encoder model and a decoder model. The encoder converts the input to a low-dimensional vector (code). The decoder reconstructs the input from this code. Typically, the encoder and decoder are both neural network models, and the entire autoencoder model is trained end-to-end. The learned code should contain sufficient information to reconstruct the inputs and can be used as input to other machine learning methods, or the autoencoder itself may be used as a generative model. }
	\label{fig:auto}
\end{figure}

Unlike discriminative models that learn $p(y|x)$ in order to predict labels $y$ given inputs $x$, generative models learn to generate examples that are not in the training set by learning the generating distribution $p(x)$ for the training data. Tantalizingly, generative models in other fields have been trained to generate new faces\cite{radford2015unsupervised}, sketches\cite{ha2017neural}, and even music\cite{roberts2018hierarchical}. Instead of using neural network models to directly learn the mapping from protein sequence to function, Sinai \textit{et al.} and Riesselman \textit{et al.} trained variational autoencoders to learn the distribution of allowed mutations within functional protein families\cite{sinai2017variational, riesselman2017deep_note}. An autoencoder is a neural network that learns to encode an input as a vector (encoding) and then reconstructs the input from the vector (decoding) (Figure~\ref{fig:auto}). By learning an encoding with smaller dimensionality than the original input, the model extracts the most important information from the input in an information-dense format. The encoding can then be used as an input to other learning algorithms. In a variational autoencoder, the learned encoding is further constrained to encourage the encodings to be densely packed to allow interpolation between examples and the ability to mix and match properties\cite{kingma2013auto}. Applied to protein sequences, variational autoencoders can learn complex epistatic relationships among variants, allowing semi-supervised predictions of variant functionality based only on existing sequences without a need for individual measurements. 

In addition, the protein variants generated by a variational autoencoder or other generative model can be highly sequence-divergent from known sequences but potentially still functional\cite{costello2019hallucinate}. These can be starting points for further engineering, or the generative model itself can be directly tuned \textit{in silico} to produce sequences optimized for a desired property. Recently, recurrent neural networks and generative adversarial networks\cite{goodfellow2014generative} have been used to generate novel antimicrobial peptides\cite{muller2018recurrent,gupta2018feedback} and protein structures\cite{anand2018generative}, and there has been an effort to develop a mathematical framework for adapting a generative model to sample sequences with one or more specified properties\cite{brookes2018design,brookes2019conditioning}. While these early examples show the potential of generative models to discover sequences with novel desired functions, this remains a promising and largely unexplored field.

Machine-learning methods have already expanded the proteins and properties that can be engineered by directed evolution. However, advances in both computational and experimental techniques, including generative models and deep mutational scanning, will also allow for better understanding of fitness landscapes and protein diversity. As researchers continue to collect sequence-function data in engineering experiments and to catalog the natural diversity of proteins, machine learning will be an invaluable tool to extract knowledge from protein data and engineer proteins for novel functions. 

\section*{Acknowledgments}

The authors wish to thank Yuxin Chen, Kadina Johnston, Bruce Wittmann, and Hopen Yang for comments on early versions of the manuscript and members of the Arnold lab, Justin Bois, and Yisong Yue for general advice and discussions on protein engineering and machine learning.

This work was supported by the U.S. Army Research Office Institute for Collaborative Biotechnologies [W911F-09-0001 to F.H.A.], the Donna and Benjamin M. Rosen Bioengineering Center [to K.K.Y.], and the National Science Foundation [GRF2017227007 to Z.W.].

The authors declare that they have no
competing financial interests.

Correspondence and requests for materials
should be addressed to F.H.A.\\
(email: frances@cheme.caltech.edu).

\bibliographystyle{naturemag}
\bibliography{master}

\end{document}